  \documentclass[aip]{revtex4-1}
\usepackage{graphicx,amsmath}\usepackage{color}
\usepackage{epstopdf}
\usepackage{epsfig}
\draft 

\begin{document}

\title{Algebraic solutions of shape-invariant position-dependent effective mass systems} 



\author{Naila Amir}
\email[]{naila.amir@live.com, naila.amir@seecs.edu.pk}
\affiliation{School of Electrical Engineering and Computer Sciences, National University of Sciences and
             Technology, Islamabad, Pakistan.}
\author{Shahid Iqbal}
\email[]{sic80@hotmail.com, siqbal@sns.nust.edu.pk}
\affiliation{School of Natural Sciences, National University of Sciences and
             Technology, Islamabad, Pakistan.}


\date{\today}

\begin{abstract}
Keeping in view the ordering ambiguity that arises due to the presence of position-dependent effective mass in the kinetic energy term of the Hamiltonian, a general scheme for obtaining algebraic solutions of quantum mechanical systems with position-dependent effective mass is discussed. We quantize the Hamiltonian of the pertaining system by using symmetric ordering of the operators concerning momentum and the spatially varying mass, initially proposed by von Roos and L\'{e}vy-Leblond. The algebraic method, used to obtain the solutions, is based on the concepts of supersymmetric quantum mechanics and shape invariance. In order to exemplify the general formalism a class of non-linear oscillators has been considered. This class includes the particular example of a one-dimensional oscillator with different position-dependent effective mass profiles. Explicit expressions for the eigenenergies and eigenfunctions in terms of generalized Hermite polynomials are presented. Moreover, properties of these modified Hermite polynomials, like existence of generating function and recurrence relations among the polynomials have also been studied. Furthermore, it has been shown that in the harmonic limit, all the results for the linear harmonic oscillator are recovered.
\end{abstract}


\maketitle 

\section{Introduction}
\noindent From last few decades quantum mechanical systems with a position-dependent effective mass (PDEM) have received considerable attention of researchers \cite{geller1993quantum,serra1997spin,morrow1984model,puente1994dipole,barranco1997structure,midya2012effect,nse,amir2014exact,amir2015coherent,amir2015ladder,bgcs2016,gcs2016}. This is due to the fact that such systems are relevant in describing many
physical situations of interest. In non-relativistic scenarios PDEM appears in many microstructures, such as compositionally graded crystals \cite{geller1993quantum}, quantum dots \cite{serra1997spin}, semiconductor heterostructure \cite{morrow1984model}, metal clusters \cite{puente1994dipole} and Helium clusters \cite{barranco1997structure} etc. The concept of PDEM arises from the effective-mass approximation which is useful in studying the motion of electrons in crystals \cite{wannier1937structure,slater1949electrons}. Recent interest in this field arose from remarkable developments in crystal-growth techniques such as molecular beam epitaxy, which allows the fabrication of non-uniform semiconductor specimens with abrupt heterojunctions \cite{bastard2008wave}, where the effective mass of the charge carriers may depend on position. \\
\noindent To study the transport of charge carriers through such heterostructures, the Schr\"{o}dinger equation can be solved either analytically or numerically by using different computational techniques. However, over the last few years a relativistic treatment has been proposed for evaluating the transport properties in condensed matter, such as the relativistic effects in the case of electron tunnelling through a multi-barrier system \cite{roy1992relativistic}, even if the relativistic effects are very small. Furthermore, this concept has been generalized to PDEM systems \cite{roy1993relativistic,cotuaescu2007applying,yannouleas2015transport}. For example, in order to avoid some difficulties of the non-relativistic theory, the Dirac equation has been used to describe quantum mechanical systems with PDEM \cite{roy1993relativistic} and a relativistic transfer matrix has been derived for a Dirac electron moving in a fixed direction through rectangular barriers of arbitrary shape \cite{cotuaescu2007applying}. Most recently, the PDEM  systems have been used to describe the transport, Aharonov-Bohm, and topological effects in graphene molecular junctions and graphene nanorings \cite{yannouleas2015transport}. \\
\noindent Due to the above mentioned and abundant other applications, interest has been developed in finding exact solutions to PDEM systems \cite{amir2014exact,amir2015ladder,carinena2007quantum,midya2009generalized,quesne2004deformed,dekar1998exactly,gonul2002exact,yu2004series,alhaidari2002solutions}. \noindent However, the study of systems with position-dependent mass distribution involves some conceptual and mathematical problems of fundamental nature. An important issue that arises in this context is to deal with the incompatible nature of the operators concerning mass and momentum in the kinetic energy term, that arises due to dependence of mass on position. The most general form of the Hamiltonian of a PDEM system, proposed by Oldwig von Roos \cite{von1983position}, is given as
\begin{equation}\label{1}
 \hat{H} = \frac{-1}{4}\bigg[m^{a}(x)\frac{d}{dx}m^{b}(x)\frac{d}{dx}m^{c}(x)+
m^{c}(x)\frac{d}{dx}m^{b}(x)\frac{d}{dx}m^{a}(x)\bigg]+V(x),
\end{equation}
where $V(x)$ represents the potential energy term of the given system and $a,b,c$ are the ambiguity parameters related by the constraint $a + b + c = -1$. It is important to remark that different choices of the ambiguity parameters, $a,~b$ and $c$, lead to distinct non-equivalent quantum Hamiltonians \cite{levy1995position,oan}. However, a particular set of values, $a=c=0$ and $b=-1$, initially suggested by Lévy-Leblond. \cite{levy1995position}, leads to symmetric ordering of the operators concerning the momentum and the position-dependent effective mass. This technique provides us with a self-adjoint quantum Hamiltonian  \cite{amir2014exact,amir2015coherent,amir2015ladder,levy1995position}
\begin{equation}\label{1.2}
 \hat{H}=
  -\bigg(\frac{1}{2m(x)}\bigg)\frac{d^{2}}{dx^{2}}-\bigg(\frac{1}{2m(x)}\bigg)^{'}\frac{d}{dx}+V(x),
\end{equation}
in the Hilbert space $L^{2}(R)$, where prime denotes the differentiation with respect to the position variable $x$.   \\
\noindent After quantizing the classical Hamiltonian one can proceed for finding the exact solutions of the given system. The traditional way of obtaining exact solutions is to solve the corresponding Schr\"{o}dinger equation for the underlying PDEM systems. However, there exist various other methods that can be more advantageous over the traditional way of solving this second order differential equation with spatially varying mass. These methods include supersymmetric quantum mechanics (SUSY QM) along with the property of shape invariance \cite{amir2015ladder,milanovic1999generation,bagchi2005deformed,plastino1999supersymmetric,gonul2002supersymmetric,samani2003shape,ganguly2007shape}, method of point canonical transformations \cite{tezcan2007exact,mustafa2006d,quesne2009point}, potential algebras \cite{amir2015ladder,kamran1990lie,roy2005lie} and path integration which relates the constant mass Green's function to that of position-dependent mass \cite{chetouani1995green,mandal2000path}. \\
\noindent The factorization method for the PDEM systems not only provides us with a powerful tool for obtaining solutions but it also enables us to determine the appropriate ladder operators for the system under consideration \cite{amir2015ladder}. This in turn allows us to obtain the underlying algebraic structure of a given system with spatially varying mass. The exact solutions and the underlying algebra of a PDEM system has vast applications in different areas of mathematics and physics, such as they play an important role in the theory of coherent states \cite{nse,amir2015coherent,ghosh2012coherent,bgcs2016,gcs2016}. Coherent states are extremely useful in various areas such as quantum mechanics \cite{iqbal2010space,iqbal2013gazeau}, quantum optics \cite{iqbal2011generalized,iqbal2012comment}, quantum information \cite{ali2013coherent} and group theory \cite{barut1971new} and have attracted attention of many researchers. In the PDEM scenario authors have made several contributions \cite{nse,amir2015coherent,ghosh2012coherent,y2009position,y20111}.\\
\noindent In our work we follow an algebraic technique to solve the quantum mechanical systems with PDEM. The history of algebraic methods goes back to the seminal work of Schr\"{o}dinger \cite{schrodinger1940method,schrodinger1940further,schrodinger1941factorization} and Infeld and Hull \cite{infeld1951factorization}. Later on these algebraic approaches were extended by using the concepts of supersymmetry \cite{witten1981dynamical,cooper1983aspects} and shape invariance \cite{gendenshtein1983derivation}, which have been extensively used to solve many quantum mechanical systems with constant mass \cite{cooper1995supersymmetry,cooper2001supersymmetry,dutt1988supersymmetry,levai1989search}. The application of supersymmetry to the non-relativistic quantum mechanics provides us with a deeper understanding of the analytically solvable Hamiltonians as well as a set of powerful approximation techniques for dealing with the systems admitting no exact solutions \cite{cooper1995supersymmetry,cooper2001supersymmetry}. A fundamental role has been played by the concept of shape invariance in these developments \cite{dutt1988supersymmetry,levai1989search,balantekin1998algebraic}.\\
\noindent Later on, this algebraic technique has been extended to incorporate position dependence of mass \cite{amir2015ladder,bagchi2005deformed,plastino1999supersymmetric,samani2003shape,gonul2002supersymmetric}. However, in the case of PDEM systems, this algebraic approach needs to be modified in order to incorporate the position dependence of mass in the kinetic energy term and related mathematical and physical difficulties. For instance, the intertwining operators are defined differently which leads to a modified shape invariance condition \cite{quesne2009point,bagchi2005deformed}. It is important to note that the earlier work in this regard is mainly focused on either the construction of the shape invariant potentials \cite{samani2003shape} or the factorization of PDEM Hamiltonians and obtaining corresponding solutions which are restricted to harmonic or Morse like spectra  \cite{plastino1999supersymmetric}. Moreover, the problem of finding the wavefunctions has been restricted
either to a few lower excited states or to the eigenfunctions obtained formally by the solutions of the
corresponding constant mass Schr\"{o}dinger equation.\\
\noindent The present work provides a generalized formalism, based on SUSY QM and shape invariance, for obtaining the algebraic solutions of quantum mechanical systems with spatially varying mass. This algebraic formalism enable us to determine the complete energy spectrum of a given PDEM system. For the sake of illustration a class of non-linear oscillators with spatially varying mass has been considered. The speciality of this particular choice is that the exact solutions for all the non-linear oscillators can be obtained by applying various approaches discussed above and the results are in complete agreement with each other \cite{amir2014exact,amir2015ladder}. In each case, it is shown that as the variable mass approaches a constant mass, all the obtained results reduce to the well known results of the linear harmonic oscillator with constant mass.\\
\noindent The organization of the paper is as follows: Section 2 is dedicated to a self-contained study of the algebraic technique for the PDEM quantum mechanical systems. In section 3, this formalism has been applied to a class of non-linear oscillators with spatially varying mass. Explicit expression for the energy spectrum and the corresponding wave functions are obtained in terms of $\lambda-$dependent modified Hermite polynomials. Properties of the family of orthogonal polynomials are also discussed. Appropriate generating functions have been introduced and recurrence relations among different orders of these modified polynomials are also provided. Finally in section 4, we close our work by some concluding remarks.
\section{Factorization method for quantum systems with position-dependent effective mass}
\noindent In this section, we present a self-contained introduction to the factorization method, based on the idea of supersymmetry and shape invariance and discuss how to determine the spectrum and the corresponding eigensates for the quantum systems with position-dependent effective mass.
\subsection{Supersymmetric quantum mechanics}
\noindent In order to apply the SUSY QM formalism, we introduce a pair of intertwining operators
\begin{eqnarray}\label{op}
  \hat{A}&=&\frac{1}{\sqrt{2m(x)}}\frac{d}{dx}+W(x), \nonumber\\
  \hat{A^{\dag}}&=&\frac{-d}{dx}\bigg(\frac{1}{\sqrt{2m(x)}}\bigg)+W(x),
\end{eqnarray}
in such a way that
\begin{equation}\label{2.2}
   \hat{H}_{-}=\frac{-1}{2m(x)}\frac{d^{2}}{dx^{2}}-\bigg(\frac{1}{2m(x)}\bigg)^{'}\frac{d}{dx}+V_{-}(x),
\end{equation}
admits the factorization
\begin{eqnarray}\label{2.3}
   \hat{H}_{-}&=&\hat{A}^{\dag}\hat{A}, \nonumber \\
              &=&\frac{-1}{2m(x)}\frac{d^{2}}{dx^{2}}-\bigg(\frac{1}{2m(x)}\bigg)^{'}\frac{d}{dx}-
                 \bigg(\frac{W}{\sqrt{2m(x)}}\bigg)^{'}+W^{2},
\end{eqnarray}
which leads to the following relation
\begin{equation}\label{c}
    \hat{H}_{-}=\hat{H}-E_{0},
\end{equation}
where $E_{0}$ is the ground-state energy. This is possible if and only if the super-potential $W(x)$, for the confining PDEM system, satisfies the following Riccati equation
\begin{equation}\label{2.3'}
   V_{-}(x)= W^{2}(x) - \bigg(\frac{W(x)}{\sqrt{2m(x)}}\bigg)^{'},
\end{equation}
where $V_{-}(x)$ is related to the potential of the original Hamilton $\hat{H}$ as
\begin{equation}\label{2.4}
V_{-}(x)=V(x)-E_{0}.
\end{equation}
It is important to note that the construction of the intertwining operators (\ref{op}) is based on the fact that they satisfy the following relation
\begin{equation}\label{cg}
\hat{A}|\varphi_{0}\rangle= 0,
\end{equation}
where $|\varphi_{0}\rangle$ is the ground-state of the system and the corresponding wave function is defined as
\begin{equation}\label{gs}
  \varphi_{0}(x) = \exp\bigg(-\int \sqrt{2 m(x)}~W(x)dx\bigg).
\end{equation}
Hence, the relation between the super-potential and the ground state wave function, is given as
\begin{equation*}
  W(x)= \frac{-\varphi_{0}'(x)}{\sqrt{2m(x)}\varphi_{0}(x)}.
\end{equation*}
Moreover, from (\ref{c}), it is clear that
\begin{equation}\label{gsh-}
\hat{H}_{-}|\varphi_{0}\rangle= 0.
\end{equation}
Thus, we may say that $|\varphi_{0}\rangle=|\varphi_{0}^{(-)}\rangle$ acts as the ground state of $\hat{H}_{-}$ with $E_{0}^{(-)}=0$.
The supersymmetric partner hamiltonian of $\hat{H_{-}}$ is defined as
\begin{eqnarray}\label{2.5}
   \hat{H}_{+}&=&\hat{A}\hat{A}^{\dag}, \nonumber \\
              &=&\frac{-1}{2m(x)}\frac{d^{2}}{dx^{2}}-\bigg(\frac{1}{2m(x)}\bigg)^{'}\frac{d}{dx}+V_{+}(x),
\end{eqnarray}
where $V_{+}(x)$ is associated partner potential, given as
\begin{equation}\label{2.6}
 V_{+}(x)= W^{2} - \bigg(\frac{W}{\sqrt{2m(x)}}\bigg)^{'}  +\frac{2W^{'}}{\sqrt{2m(x)}}-\bigg(\frac{1}{\sqrt{2m(x)}}\bigg)\bigg(\frac{1}{\sqrt{2m(x)}}\bigg)^{''} .
\end{equation}
The above equation may be rewritten as
\begin{equation*}
V_{+}(x)=V_{-}(x)+\frac{2W^{'}}{\sqrt{2m(x)}}-\bigg(\frac{1}{\sqrt{2m(x)}}\bigg)\bigg(\frac{1}{\sqrt{2m(x)}}\bigg)^{''}.
\end{equation*}
Note that the potential $V_{+}(x)$ depends not only on the form of the variable mass $m(x)$ but also on its supersymmetric partner potential $V_{-}(x)$.
In order to examine the underlying supersymmetry of this formalism, we introduce the super charges
\begin{equation}\label{sc}
  \hat{Q}=\left(
            \begin{array}{cc}
              0 & 0 \\
              \hat{A} & 0 \\
            \end{array}
          \right),~~~~
  \hat{Q}^{\dag}=\left(
            \begin{array}{cc}
              0 & \hat{A}^{\dag} \\
              0 & 0 \\
            \end{array}
          \right),
\end{equation}
that satisfy the SUSY QM superalgebra
\begin{equation}\label{sa}
 \{\hat{Q},\hat{Q}\} = 0,~~
 \{\hat{Q}^{\dag},\hat{Q}^{\dag}\} = 0,~~
 \{\hat{Q},\hat{Q}^{\dag}\} = \hat{H},
\end{equation}
where $\hat{H}=\left(
            \begin{array}{cc}
              \hat{H}_{-} & 0 \\
              0 & \hat{H}_{+} \\
            \end{array}
          \right)$.
Moreover, the partner Hamiltonians $\hat{H_{-}}$ and $\hat{H_{+}}$ are intertwined i.e., $\hat{A}\hat{H}_{-}=\hat{H}_{+}\hat{A}$ and $\hat{A}^{\dag}\hat{H}_{+}=\hat{H}_{-}\hat{A}^{\dag}$. This indicates that there exist a relationship among their eigenenergies and eigenstates as
\begin{eqnarray}\label{rew}
  E_{n}^{(+)}&=&E_{n+1}^{(-)},\nonumber\\
|\varphi_{n}^{(+)}\rangle&=&[E_{n}^{(+)}]^{-1/2}\hat{A}|\varphi_{n+1}^{(-)}\rangle,\\
\nonumber|\varphi_{n+1}^{(-)}\rangle&=&[E_{n}^{(+)}]^{-1/2}\hat{A}^{\dag}|\varphi_{n}^{(+)}\rangle,~~n=0,1,2,....
\end{eqnarray}
If the eigenvalues and the eigenfunctions of $\hat{H}_{-}$ are known, one can immediately solve for the
eigenvalues and the eigenfunctions of Hamiltonian $\hat{H}_{+}$. These relationships, however, do not guarantee the solvability of either of the partner potentials $V_{\pm}(x)$. In principle, one would need to have found the solutions of one of the partner Hamiltonians by some standard method, in order to obtain the solutions for the other. However, if these Hamiltonians have the additional property of shape invariance, one can determine all eigenvalues and eigenfunctions of both
partners without solving their Schr\"{o}dinger equations, as traditional methods requires.
\subsection{Shape invariance}
\noindent If both of the partner Hamiltonians depend on a parameter $\alpha$ and are related to each other in such a way that their associated potentials have the same functional form but for different value of the parameter $\alpha$, then the isospectral Hamiltonians are said to be shape invariant. This means that there exist a function $f$ such that $\alpha_{n+1}=f(\alpha_{n})$ and
\begin{equation}\label{2.7}
\hat{H_{+}}(\alpha_{n})=\hat{H_{-}}(\alpha_{n+1})+R(\alpha_{n}).
\end{equation}
Here $R(\alpha_{n})$ is the remainder term independent of the dynamical variables. The replacement of the set of parameters $\alpha_{n}$ with the $\alpha_{n+1}$ in Eq. (\ref{2.7}), is achieved by means of a similarity transformation
\begin{equation*}
  T(\alpha_{n})\hat{O}(\alpha_{n})T^{-1}(\alpha_{n})=\hat{O}(\alpha_{n+1}),
\end{equation*}
where $T(\alpha_{n})$ is an operator denoting the reparametrization,
\begin{equation*}
  T\varphi(x,\alpha_{n})=\varphi(x,\alpha_{n+1}).
\end{equation*}
The most commonly discussed classes of shape invariant potentials are:
\begin{enumerate}
  \item Translational shape invariance:  $\alpha_{n+1}=f(\alpha_{n})=\alpha_{n}+\eta$.
  \item Scaling shape invariance:  $\alpha_{n+1}=f(\alpha_{n})=\eta\alpha_{n},~~0<\eta<1$.
  \item Cyclic shape invariance:  $\alpha_{n+k}=f^{k}(\alpha_{n})=\alpha_{n}$.
\end{enumerate}
However, in the present work we will restrict ourselves to the first class only, i.e., we are only interested in translational shape invariant systems with position-dependent effective mass which are known to be exactly solvable.
\subsection{Determination of eigenenergies and eigenfunctions}
\noindent The significance of the shape invariance condition is that it allows us to determine the complete spectrum of the underlying system without even referring to its Schr\"{o}dinger equation \cite{gendenshtein1983derivation,carinena2004one,dutt1988supersymmetry}. Since the partner Hamiltonians $\hat{H_{+}}(\alpha_{n})$ and $\hat{H_{-}}(\alpha_{n+1})$, differ only by a constant, they share common eigenfunctions, and their eigenvalues are related by the same additive constants as the Hamiltonians themselves, i.e.,
\begin{eqnarray}\label{nc}
\nonumber |\varphi_{n}^{(+)}(\alpha_{n})\rangle&=&|\varphi_{n}^{(-)}(\alpha_{n+1})\rangle,\\
E_{n}^{(+)}(\alpha_{n})&=&E_{n}^{(-)}(\alpha_{n+1})+R(\alpha_{n}).
\end{eqnarray}
As already mentioned that $|\varphi_{0}^{(-)}(\alpha_{n})\rangle=|\varphi_{0}(\alpha_{n})\rangle$ is the ground state energy of $\hat{H}_{-}(\alpha_{n})$ with zero energy. First relation of (\ref{nc}) suggests that
\begin{equation*}
|\varphi_{0}^{(+)}(\alpha_{1})\rangle=|\varphi_{0}^{(-)}(\alpha_{2})\rangle=|\varphi_{0}(\alpha_{2})\rangle.
\end{equation*}
Let us now determine the excited states and corresponding eigenenergies of $\hat{H}_{-}$.
Using the intertwining relation $\hat{H}_{-}(\alpha_{1})\hat{A}^{\dag}(\alpha_{1})=\hat{A}^{\dag}(\alpha_{1})\hat{H}_{+}(\alpha_{1})$, together with integrability (\ref{2.7}), we get
\begin{eqnarray*}
\hat{H}_{-}(\alpha_{1})[\hat{A}^{\dag}(\alpha_{1})|\varphi_{0}(\alpha_{2})\rangle]&=&
\hat{A}^{\dag}(\alpha_{1})\hat{H}_{+}(\alpha_{1})|\varphi_{0}(\alpha_{2})\rangle,\nonumber \\
&=&\hat{A}^{\dag}(\alpha_{1})[\hat{H}_{-}(\alpha_{2})+R(\alpha_{1})]|\varphi_{0}(\alpha_{2})\rangle, \nonumber \\
&=&R(\alpha_{1})[\hat{A}^{\dag}(\alpha_{1})|\varphi_{0}(\alpha_{2})\rangle],
\end{eqnarray*}
suggesting that $\hat{A}^{\dag}(\alpha_{1})|\varphi_{0}(\alpha_{2})\rangle$ is an excited state of $\hat{H}_{-}$, with eigenenergy $E^{(-)}_{1}=R(\alpha_{1})$. Moreover,
\begin{eqnarray*}
  \hat{H}_{+}(\alpha_{1})|\varphi_{0}(\alpha_{2})\rangle,
  &=&\hat{H_{-}}(\alpha_{2})|\varphi_{0}(\alpha_{2})\rangle+R(\alpha_{1})|\varphi_{0}(\alpha_{2})\rangle\\
  &=&R(\alpha_{1})|\varphi_{0}(\alpha_{2})\rangle.
\end{eqnarray*}
This means that $|\varphi_{0}(\alpha_{2})\rangle$ is an eigenstate of $\hat{H}_{+}$ with energy $E_{0}^{(+)}=R(\alpha_{1})$. Iteration of above process, provides us with the eigenenergies of $\hat{H}_{-}$, as
\begin{equation}\label{2.11}
  E^{(-)}_{n} = \sum_{i=1}^{n}R(\alpha_{i}),~~E^{(-)}_{0}=0.
\end{equation}
By using the last relation the energy spectrum of the Hamiltonian $\hat{H}$ defined in Eq. (\ref{1.2}), come out to be
\begin{equation}\label{2.11'}
 E_{n}= E^{(-)}_{n} + E_{0},
\end{equation}
where $E^{(-)}_{n}$ are the eigenenergies of the $\hat{H}_{-}$ and $E_{0}$ is the ground state energy of $\hat{H}$. The corresponding eigenstates of the Hamiltonian $\hat{H}$ are given as
\begin{equation}\label{2.12}
|\varphi_{n}(\alpha_{1})\rangle \propto \hat{A}^{\dag}(\alpha_{1})\hat{A}^{\dag}(\alpha_{2})...\hat{A}^{\dag}(\alpha_{n})|\varphi_{0}(\alpha_{n+1})\rangle.
\end{equation}
Note that the above relation for the wave functions is a generalization of the algebraic method of obtaining the energy eigenstates for the standard one-dimensional harmonic oscillator with constant mass. For the sake of convenience, it is often better to have an explicit expression for these eigenstates, so instead of using the above relation (\ref{2.12}), the following simplified form can be used \cite{dutt1988supersymmetry}
\begin{equation}\label{2.13}
  |\varphi_{n}(\alpha_{1})\rangle = \hat{A}^{\dag}(\alpha_{1})|\varphi_{n-1}(\alpha_{2})\rangle.
\end{equation}
Thus, we conclude that SUSY QM along with the property of shape invariance provides us
with an excellent tool to determine the entire spectrum of solvable quantum systems through
a step-by-step algebraic procedure, without going into the details of solving the corresponding
Schr\"{o}dinger equation. Also, it is important to note that the shape invariance condition does
not always help one in determining the spectrum. Another important ingredient required in
this regard is the unbroken supersymmetry \cite{cooper1983aspects,cooper2001supersymmetry}. In the ongoing analysis, we will assume that the supersymmetry is unbroken.
\section{Class of non-linear harmonic oscillators with PDEM}
\noindent Many realistic phenomena in nature exhibit nonlinear oscillations which have motivated researchers to explore non-linear oscillators. In order to exemplify the general procedure discussed in the previous sections we consider a class of non-linear oscillators with PDEM. All these systems are exactly solvable possessing translational shape-invariant potentials. We discuss the first example in detail and remaining can be solved following the same procedure as adopted for the first one.\\
Here it is important to emphasize that in all of the examples considered below we have used symmetric ordering approach \cite{von1983position,levy1995position} in order to quantize the corresponding Hamiltonians, as mentioned before in the general construction.
\subsection{Case 1: $m(x)=(1+\lambda x^{2})^{-1}$}
\begin{figure}
\centering
\includegraphics[width=0.90\textwidth]{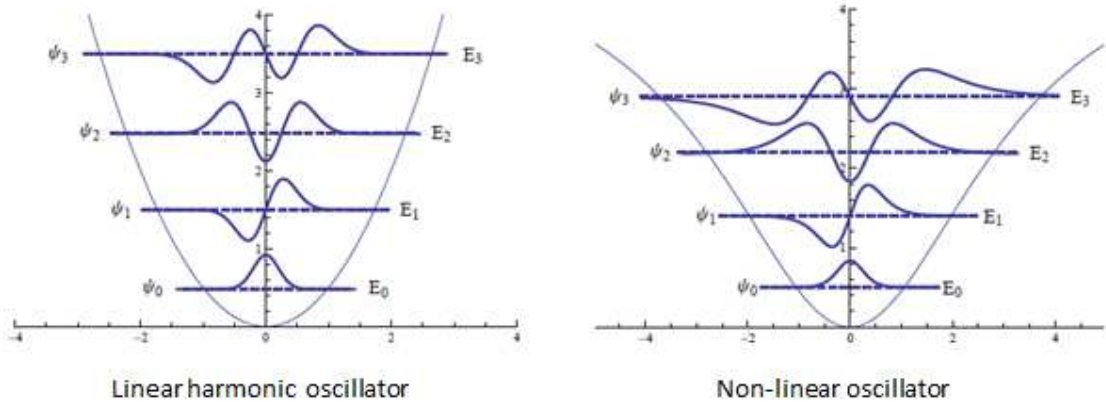}
\caption[Comparison between standard harmonic oscillator and non-linear oscillator with PDEM.]{Comparison between standard harmonic oscillator and non-linear oscillator with PDEM for $\tilde{\lambda}=0.1$.} \label{cln}
\end{figure}

\noindent As a first example, let us examine the non-linear harmonic oscillator with PDEM \cite{carinena2007quantum,midya2009generalized}, which was initially considered by Mathews and Lakshmanan \cite{mathews1975quantum,lakshmanan2003nonlinear}. They studied a non-linear differential equation of the form
\begin{equation}\label{1.3}
  (1+\lambda x^{2})\ddot{x}-(\lambda x)\dot{x}^{2}+\alpha^{2}x=0.
\end{equation}
It was proved that solution of this equation represents non-linear oscillations with quasi-harmonic form. Also it was shown that such system is described by the Lagrangian
\begin{equation}\label{n3.3}
  L=\frac{1}{2}\bigg[\frac{1}{1+\lambda x^{2}}\bigg](\dot{x}-\alpha^{2}x^{2}).
\end{equation}
Thus, the non-linear system can be considered as an harmonic oscillator with PDEM given by
\begin{equation}\label{n3.4}
  m(x)=\frac{1}{1+\lambda x^{2}}.
\end{equation}
The Lagrangian introduced in Eq. (\ref{n3.3}) and the spatially varying mass given in Eq. (\ref{n3.4}), give rise to the momentum of the non-linear harmonic oscillator that can be written as
\begin{equation}\label{n3.5}
 p=\frac{\partial L}{\partial\dot{x}}=\frac{\dot{x}}{1+\lambda x^{2}},
\end{equation}
which enables us to write the classical Hamiltonian of the given system as
\begin{equation}\label{ch}
    H=\frac{1}{2}(1+\lambda x^{2})p^{2}+
    \frac{1}{2}\bigg(\frac{\alpha^{2}x^{2}}{1+\lambda x^{2}}\bigg),
\end{equation}
where $\lambda$ represents the non-linearity parameter which can be positive as well as negative. Note that, for negative values of $\lambda$, there exists a singularity for the given mass function and associated dynamics, at $1-|\lambda|x^{2} = 0$. Therefore, for $\lambda<0$, our analysis is restricted to the interior of the interval $x^{2}\leq1/|\lambda|$.
For the sake of quantization of the classical Hamiltonian defined in Eq. (\ref{ch}), we make use of Eq. (\ref{1.2}), and get the a quantum Hamiltonian of the form
\begin{equation}\label{h}
\hat{H}= \frac{1}{2}\left[
                         \begin{array}{c}
                          -(1+\lambda x^{2})\frac{d^{2}}{dx^{2}}-2\lambda x\frac{d}{dx}
                          +\frac{\alpha^{2}x^{2}}{1+\lambda x^{2}}  \\
                         \end{array}
                       \right],
\end{equation}
which is manifestly Hermitian in the space $L^{2}(R)$ for $\lambda >0$ and $L^{2}[-1/\sqrt{|\lambda|},1/\sqrt{|\lambda|}]$ for negative values of $\lambda$.\\
As a general exposition, we mention here that the non-linear oscillator (\ref{ch}) has been considered by various authors \cite{carinena2004non,carinena2007quantum,carinena2004one,carinena2008quantization}. In order to obtain the quantum Hamiltonian, these authors have used a different quantization scheme, based on the theory of symmetries that make use of the existence of Killing vectors and invariant measure, which suggests that, the quantum Hamiltonian is self adjoint in the space $L^{2}(R, d\mu)$, where $d\mu = (1 + \lambda x^{2})^{\frac{-1}{2}}dx$, instead of the standard space $L^{2}(R)$. \\
In order to apply the SUSY QM formalism, we introduce a pair of intertwining operators given in Eq. (\ref{op}), as
\begin{eqnarray}\label{3.2}
  \hat{A} &=& \frac{1}{\sqrt{2}}\bigg[\sqrt{1+\lambda x^{2}}\frac{d}{dx}+\frac{\alpha x}{\sqrt{1+\lambda x^{2}}}\bigg], \nonumber \\
  \hat{A}^{\dag} &=& \frac{1}{\sqrt{2}}\bigg[-\frac{d}{dx}(\sqrt{1+\lambda x^{2}})+\frac{\alpha x}{\sqrt{1+\lambda x^{2}}}\bigg]  
                 = \frac{1}{\sqrt{2}}\bigg[-\sqrt{1+\lambda x^{2}}\frac{d}{dx}+\frac{(\alpha - \lambda) x}{\sqrt{1+\lambda x^{2}}}\bigg],
\end{eqnarray}
such that the condition introduced in Eq. (\ref{cg}), is satisfied. Note that this is just a first order differential equation whose solutions provides us with
\begin{equation}\label{3.11}
  \varphi_{0}(x)=\mathcal{N}_{0}(1+\lambda x^{2})^{\frac{-\alpha}{2\lambda}},
\end{equation}
where $\mathcal{N}_{0}$ is the normalization constant. By means of the intertwining operators introduced in Eq. (\ref{3.2}), we obtain the supersymmetric Hamiltonian $\hat{H}_{-}$ given in Eq. (\ref{2.2}), as
\begin{equation}\label{3.3}
 \hat{H}_{-}=\hat{A}^{\dag}\hat{A}=\frac{1}{2}\left[
                         \begin{array}{c}
                          -(1+\lambda x^{2})\frac{d^{2}}{dx^{2}}-2\lambda x\frac{d}{dx}  \\
                         \end{array}
                       \right]+V_{-}(x),
\end{equation}
where the corresponding potential $V_{-}(x)$ defined in Eq. (\ref{2.3'}), as
\begin{equation}\label{v1}
  V_{-}(x)=\frac{1}{2}\frac{\alpha^{2}x^{2}}{1+\lambda x^{2}}-\frac{\alpha}{2}.
\end{equation}
Comparison of Eqs. (\ref{h}) and (\ref{3.3}), yields
\begin{equation}\label{3.7}
  \hat{H}=\hat{H}_{-}+\frac{\alpha}{2},
\end{equation}
where $E_{0}=\alpha/2$ acts as the ground state energy for the Hamiltonian $\hat{H}$.\\
\noindent The supersymmetric partner Hamiltonian $\hat{H}_{+}$ defined in Eq. (\ref{2.5}), can be written as
\begin{equation}\label{3.4}
 \hat{H}_{+}=\hat{A}\hat{A}^{\dag}=\frac{1}{2}\left[
                         \begin{array}{c}
                          -(1+\lambda x^{2})\frac{d^{2}}{dx^{2}}-2\lambda x\frac{d}{dx} \\
                         \end{array}
                       \right]+V_{+}(x),
\end{equation}
where
\begin{equation}\label{v2}
  V_{+}(x)=\frac{1}{2}\frac{(\alpha-\lambda)^{2}x^{2}}{1+\lambda x^{2}}+\frac{1}{2}(\alpha-\lambda).
\end{equation}
Note that the partner Hamiltonians $\hat{H}_{-}$ and $\hat{H}_{+}$ are related to each other by means of the following integrability condition
\begin{equation}\label{3.8}
  \hat{H}_{+}(\alpha_{1})=\hat{H}_{-}(\alpha_{2})+R(\alpha_{1}),
\end{equation}
where $\alpha_{1}=\alpha,~~\alpha_{2}=f(\alpha)=\alpha-\lambda$ and $R(\alpha_{1})=\alpha-\lambda$. In order to find the spectrum and the associated wave functions of the position-dependent oscillator, we follow the same procedure that we discussed in the previous section. As suggested in Eq. (\ref{2.11}), the energy eigenvalues for the shape invariant systems are given by
\begin{equation*}
  E^{(-)}_{n} = \sum_{i=1}^{n}R(\alpha_{i}),~~~E^{(-)}_{0}=0.
\end{equation*}
By inserting the values of $R(\alpha_{i})$ in last equation we arrive at
\begin{eqnarray}\label{3.9}
  E^{(-)}_{n} &=& \sum_{i=1}^{n}(\alpha_{i}-\lambda), \nonumber \\
        &=&n\alpha - \lambda\bigg[\frac{n(n+1)}{2}\bigg],~~~~~~~n=0,1,2,....
\end{eqnarray}
Note that $E^{(-)}_{n}$ are the eigenenergies for the Hamiltonians $\hat{H}_{-}$ given in Eq. (\ref{3.3}). Now consider Eq. (\ref{3.7}) and observe that $\hat{H}=\hat{H}_{-}+\frac{\alpha}{2}$. So, in order to get the energy spectrum for the hamiltonian $\hat{H}$ given in Eq. (\ref{h}), we will just incorporate the additional term $``\frac{\alpha}{2}"$ to the spectrum of $\hat{H}_{-}$, due to which all the energy levels will get shifted. Thus, the eigenvalues for $\hat{H}$ turns out to be
\begin{equation}\label{3.10}
  E_{n} = \alpha \bigg(n+\frac{1}{2}\bigg) - \lambda\bigg(\frac{n(n+1)}{2}\bigg),~~~~~~~n=0,1,2,....
\end{equation}
Note that for $\lambda=0$, we get back the energy eigenvalues of the usual harmonic oscillator. \\
\noindent Our next target is to obtain all the excited states explicitly. For this we make use of Eq. (\ref{2.13}) and obtain the first excited state of non-linear harmonic oscillator as
\begin{equation*}
  |\varphi_{1}(\alpha_{1})\rangle = \hat{A}^{\dag}(\alpha_{1})|\varphi_{0}(\alpha_{2})\rangle.
\end{equation*}
Substitution of Eqs. (\ref{3.2}) and (\ref{3.11}) in the last equation yields
\begin{equation}\label{3.12a}
|\varphi_{1}(\alpha_{1})\rangle =\frac{\mathcal{N}_{0}}{\sqrt{2}}
\bigg[\frac{\alpha x}{\sqrt{1+\lambda x^{2}}}-\frac{d}{dx}(\sqrt{1+\lambda x^{2}})\bigg](1+\lambda x^{2})^{\frac{-\alpha}{2\lambda}+\frac{1}{2}}.
\end{equation}
By making use of the following identity
\begin{equation}\label{3.13}
 \bigg[\frac{\alpha x}{\sqrt{1+\lambda x^{2}}}-\frac{d}{dx}(\sqrt{1+\lambda x^{2}})\bigg]h_{1}(x)=
(-1) \bigg[(1+\lambda x^{2})^{\frac{\alpha}{2\lambda}}
\frac{d}{dx}(1+\lambda x^{2})^{\frac{-\alpha}{2\lambda}+\frac{1}{2}}\bigg]h_{1}(x),
\end{equation}
where $h_{1}(x)$ is any arbitrary differential function, (\ref{3.12a}) may be rewritten in a simplified form as
\begin{equation}\label{3.12b}
 |\varphi_{1}(\alpha_{1})\rangle =(-1)\mathcal{N}_{1}
\bigg[(1+\lambda x^{2})^{\frac{\alpha}{2\lambda}}\frac{d}{dx}(1+\lambda x^{2})^{\frac{-\alpha}{\lambda}+1}\bigg].
\end{equation}
Note that, if we look at equation (\ref{3.13}) carefully, we notice that left hand side of this expression is of the same form as $\hat{A}^{\dag}$. Therefore, we may rewrite Eq. (\ref{3.13}) as
\begin{equation}\label{3.14}
\hat{A}^{\dag}=\frac{-1}{\sqrt{2}}\bigg[(1+\lambda x^{2})^{\frac{\alpha}{2\lambda}}\frac{d}{dx}(1+\lambda x^{2})^{\frac{-\alpha}{2\lambda}+\frac{1}{2}}\bigg].
\end{equation}
Using (\ref{3.11}) and (\ref{3.14}) in equation (\ref{2.13}), the energy eigenstates turn out to be
\begin{equation}\label{3.15}
\varphi_{n}=\mathcal{N}_{n}(-1)^{n}(1+\lambda x^{2})^{\frac{-\alpha}{2\lambda}}
                  \bigg[(1+\lambda x^{2})^{\frac{\alpha}{\lambda}}\frac{d^{n}}{dx^{n}}(1+\lambda x^{2})^{n}(1+\lambda x^{2})^{\frac{-\alpha}{\lambda}}\bigg],
\end{equation}
where $\mathcal{N}_{n}$ are the normalization constants.\\
\noindent By introducing the dimensionless parameters $\zeta=\sqrt{\alpha}x$ and $\lambda=\alpha \tilde{\lambda}$,
the closed form relation for eigenenergies and the corresponding wave functions, obtained in Eq. (\ref{3.10}) and Eq. (\ref{3.15})
respectively, may be rewritten as
\begin{equation}\label{3.10'}
 E_{n} = \alpha \bigg[\bigg(n+\frac{1}{2}\bigg) - \tilde{\lambda}\bigg(\frac{n(n+1)}{2}\bigg)\bigg],~~~~~~~n=0,1,2,....
\end{equation}
and
\begin{equation}\label{3.16}
 \varphi_{n} = \mathcal{N}_{n}(-1)^{n}(1+\tilde{\lambda} \zeta^{2})^{\frac{-1}{2\tilde{\lambda}}}
               \bigg[(1+\tilde{\lambda} \zeta^{2})^{\frac{1}{\tilde{\lambda}}}\frac{d^{n}}{d\zeta^{n}}(1+\tilde{\lambda} \zeta^{2})^{n}(1+\tilde{\lambda} \zeta^{2})^{\frac{-1}{\tilde{\lambda}}}\bigg],
\end{equation}
respectively. Note that
\begin{equation*}
   \lim_{\tilde{\lambda}\rightarrow 0} \bigg[(-1)^{n}(1+\tilde{\lambda} \zeta^{2})^{\frac{1}{\tilde{\lambda}}}\frac{d^{n}}{dz^{n}}(1+\tilde{\lambda} \zeta^{2})^{n-\frac{1}{\tilde{\lambda}}}\bigg]
 =(-1)^{n}e^{\zeta^{2}}\frac{d^{n}}{dz^{n}}e^{-\zeta^{2}}
 =\mathcal{H}_{n}(\zeta).
\end{equation*}
Thus it is clear from above that
\begin{equation}\label{3.17}
\mathcal{H}_{n}(\zeta,\tilde{\lambda})= (-1)^{n}\bigg[(1+\tilde{\lambda} \zeta^{2})^{\frac{1}{\tilde{\lambda}}}\frac{d^{n}}{dz^{n}}(1+\tilde{\lambda} \zeta^{2})^{n-\frac{1}{\tilde{\lambda}}}\bigg],~~~n=0,1,2,....
\end{equation}
must be considered as the \textbf{Rodrigues formula} for the modified Hermite polynomials analogous to the standard ones. Substitution of last expression in (\ref{3.16}) yields
\begin{equation}\label{3.16'}
 \varphi_{n} = \mathcal{N}_{n}\mathcal{H}_{n}(\zeta,\tilde{\lambda})
               (1+\tilde{\lambda}\zeta^{2})^{\frac{-1}{2\tilde{\lambda}}},~~~n=0,1,2,.....
\end{equation}
\begin{figure}
\centering
\includegraphics[width=0.70\textwidth]{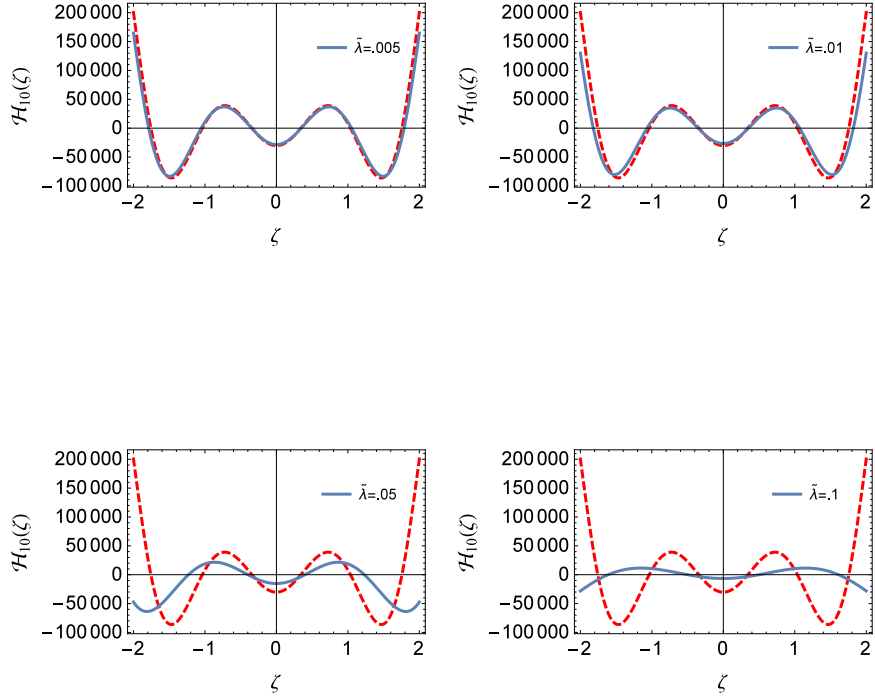}
\caption[Comparison between standard Hermite polynomials and $\tilde{\lambda}$-dependent Hermite polynomials.]{Comparison between standard Hermite polynomials (dotted line) and $\tilde{\lambda}$-dependent Hermite polynomials (solid line) for different values of $\tilde{\lambda}$.} \label{hn}
\end{figure}
The wave functions obtained are expressed in terms of the modified Hermite polynomials. So it is natural to expect that for $\tilde{\lambda}=0$ all the properties the standard Hermite polynomial are retained. The effect of the non-linearity parameter $\tilde{\lambda}$, that differentiates our modified Hermite polynomial from the standard Hermite polynomials is evident from Fig. (\ref{hn}) which shows a comparison between standard Hermite polynomials and $\tilde{\lambda}-$dependent Hermite polynomials.\\
Furthermore, comparison of linear harmonic oscillator and non-linear oscillator with PDEM can be seen easily from Fig. (\ref{cln}), which shows that the potential well in both cases is different. Moreover, it is clear from Fig. (\ref{cln}) that energy spectrum is not equispaced in the case of non-linear oscillator with PDEM in comparison with the harmonic oscillator with constant mass.\\
\noindent We now determine an appropriate generating function $g(\zeta,\tilde{\lambda},t)$ for these $\tilde{\lambda}-$dependent
Hermite polynomials such that
\begin{equation*}
  \lim_{\tilde{\lambda}\rightarrow 0} g(\zeta,\tilde{\lambda},t)=g(\zeta,t),
\end{equation*}
where $g(\zeta,t)=e^{2t\zeta-t^{2}}$ is the generating function for the standard Hermite polynomial. Let us choose the following form of the generating function for the representation of the modified Hermite polynomials
\begin{equation}\label{3.18}
  g(\zeta,\tilde{\lambda},t)=[1+\tilde{\lambda}(2t\zeta-t^{2})]^{\frac{1}{\tilde{\lambda}}-\frac{1}{2}}.
\end{equation}
Note that this choice of generating function satisfies the correct limit defined above. Due to the existence of this generating function we can obtain the recurrence relations for the modified Hermite polynomials in parallel to the standard Hermite polynomials. The generating function defined in (\ref{3.18}), in terms of power series is expressed as
\begin{equation}\label{3.19}
 [1+\tilde{\lambda}(2t\zeta-t^{2})]^{\frac{1}{\tilde{\lambda}}-\frac{1}{2}}=\sum_{m=0}^{\infty}\mathcal{H}_{m}(\zeta,\tilde{\lambda})\frac{t^{m}}{m!}.
\end{equation}
Expanding the L.H.S of the above expression and equating the coefficients of powers of $t$, we obtain explicit expressions for first few modified Hermite polynomials as
\begin{eqnarray*}
 \mathcal{H}_{0} &=& 1, \\
  \mathcal{H}_{1} &=& (2-\tilde{\lambda})\zeta, \\
 \mathcal{H}_{2} &=& (-1)(2-\tilde{\lambda})[1+(3\tilde{\lambda}-2)\zeta^{2}], \\
 \mathcal{H}_{3} &=& (-3)(2-\tilde{\lambda})(2-3\tilde{\lambda})\bigg[\zeta+\frac{1}{3}(5\tilde{\lambda}-2)\zeta^{3}\bigg],\\
 \mathcal{H}_{4} &=& (3)(2-\tilde{\lambda})(2-3\tilde{\lambda})\bigg[1+2(5\tilde{\lambda}-2)\zeta^{2}+\frac{1}{3}(5\tilde{\lambda}-2)(7\tilde{\lambda}-2)\zeta^{4}\bigg],\\
 \mathcal{H}_{5} &=& (15)(2-\tilde{\lambda})(2-3\tilde{\lambda})(2-5\tilde{\lambda})\bigg[\zeta+\frac{2}{3}(7\tilde{\lambda}-2)\zeta^{3}+\frac{1}{15}(7\tilde{\lambda}-2)
 (9\tilde{\lambda}-2)\zeta^{5}\bigg].
 \end{eqnarray*}
Note that for $\tilde{\lambda}= 0$, these expressions are exactly the same that we have in case of standard Hermite polynomials. Also it is important to remark that these relations coincides with those that we obtained by solving the system analytically \cite{amir2014exact}.\\
By taking derivative of the equation (\ref{3.19}) with respect to $``\zeta"$ and simplifying we arrive at
\begin{equation*}
  (2-\tilde{\lambda})\sum_{m=0}^{\infty}\mathcal{H}_{m}(\zeta,\tilde{\lambda})\frac{t^{m+1}}{m!}=\sum_{m=0}^{\infty}\mathcal{H}^{'}_{m}(\zeta,\tilde{\lambda})\frac{t^{m}}{m!}+2\zeta\tilde{\lambda} \sum_{m=0}^{\infty}\mathcal{H}^{'}_{m}(\zeta,\tilde{\lambda})\frac{t^{m+1}}{m!}-\tilde{\lambda} \sum_{m=0}^{\infty}\mathcal{H}^{'}_{m}(\zeta,\tilde{\lambda})\frac{t^{m+2}}{m!},
\end{equation*}
which leads to the following recursion relation
\begin{equation}\label{3.20}
  m(2-\tilde{\lambda})\mathcal{H}_{m-1}=\mathcal{H}^{'}_{m}+2\zeta\tilde{\lambda} m\mathcal{H}^{'}_{m-1}-\tilde{\lambda} m(m-1)\mathcal{H}^{'}_{m-2},~~~m=0,1,2,.....
\end{equation}
Note that for $\tilde{\lambda}\rightarrow 0$ the above relation coincides with the recursion relation for the standard Hermite polynomial i.e., we get $\mathcal{H}^{'}_{m}=2m\mathcal{H}_{m-1}$.\\
Let us now take the derivative of (\ref{3.19}) with respect to $``t"$ and simply, we finally arrive at
\begin{eqnarray*}
\zeta(2-\tilde{\lambda})\sum_{m=0}^{\infty}\mathcal{H}_{m}(\zeta,\tilde{\lambda})
\frac{t^{m}}{m!}-(2-\tilde{\lambda})\sum_{m=0}^{\infty}\mathcal{H}_{m}(\zeta,\tilde{\lambda})\frac{t^{m+1}}{m!}
=\\
\sum_{m=0}^{\infty}\mathcal{H}_{m+1}(\zeta,\tilde{\lambda})\frac{t^{m}}{m!}+2\zeta\tilde{\lambda} \sum_{m=0}^{\infty}\mathcal{H}_{m+1}(\zeta,\tilde{\lambda})\frac{t^{m+1}}{m!}
 -\tilde{\lambda} \sum_{m=0}^{\infty}\mathcal{H}_{m+1}(\zeta,\tilde{\lambda})\frac{t^{m+2}}{m!}.
\end{eqnarray*}
This leads to the following expression
\begin{equation}\label{3.21}
  \mathcal{H}_{m+1}-\zeta[(2-\tilde{\lambda})-2\tilde{\lambda} m]\mathcal{H}_{m}+m[(2-\tilde{\lambda})-\tilde{\lambda}(m-1)]\mathcal{H}_{m-1}=0,~~~m=0,1,2,...,
\end{equation}
and for $\tilde{\lambda}\rightarrow o$, we obtain the well known recurrence relation for the standard Hermite polynomial give as
\begin{equation*}
  \mathcal{H}_{m+1}-2\zeta\mathcal{H}_{m}+2m\mathcal{H}_{m-1}=0,~~~m=0,1,2,...
\end{equation*}
We have obtained the exact solutions to the non-linear harmonic oscillator with position-dependent effective mass by using supersymmetric formalism along with the property of shape invariance. It is important to remark here that these solutions are in complete agreement with the results that we have already obtained by the solving the PDEM Schr\"{o}dinger analytically \cite{amir2014exact}. Also, same solutions can be obtained by using appropriate ladder operators for the given system \cite{amir2015ladder}. Furthermore, these results also coincide with the ones that have been obtained by applying a perturbative approach to solve the system under consideration \cite{amir2015coherent}. It is worth mentioning over here that same analysis is valid for the remaining examples.
\subsection{Case 2: $m(x)=\bigg(1+\frac{x^{2}}{\lambda}\bigg)^{-1}$}
\noindent Let us consider the non-linear harmonic oscillator with spatially varying mass of the form
\begin{equation}\label{e2.1}
m(x)=\bigg(1+\frac{x^{2}}{\lambda}\bigg)^{-1},
\end{equation}
where $m(x)\rightarrow 1$ as the non-linearity parameter $\lambda\rightarrow \infty$. For this particular choice of PDEM, the quantum Hamiltonian is of the form
\begin{equation}\label{e2.2}
\hat{H}=\frac{1}{2}-\bigg(1+\frac{x^{2}}{\lambda}\bigg)\frac{d^{2}}{dx^{2}}-\frac{x}{\lambda}\frac{d}{dx}
+V(x),
\end{equation}
where
\begin{equation*}
  V(x)=\frac{1}{2}m(x)\alpha^{2}x^{2}
      =\frac{1}{2}\frac{\alpha^{2}x^{2}}{\big(1+\frac{x^{2}}{\lambda}\big)}.
\end{equation*}
Similar to the previous example, here again $\lambda$ can take both positive and negative values and for the present case our analysis is  restricted to the interval $-\sqrt{|\lambda|}<x < \sqrt{|\lambda|}$. Thus for positive values of the non-linearity $\lambda$, the quantum Hamiltonian (\ref{e2.2}), is self-adjoint in the space $L^{2}(R)$ and for $\lambda<0$, the space reduces to $L^{2}[-\sqrt{|\lambda|},\sqrt{|\lambda|}]$. \\
In order to determine the isospectral Hamiltonians $\hat{H}_{-}$ and $\hat{H}_{+}$, we consider the super potential $W(x)$ as
\begin{equation}\label{e2.3}
W(x)=\frac{1}{\sqrt{2}}\frac{\alpha x}{\sqrt{1+\frac{x^{2}}{\lambda}}},
\end{equation}
so that the pair of intertwining operators defined in Eq. (\ref{op}), are obtained as
\begin{equation}\label{e2.4'}
  \hat{A} = \frac{1}{\sqrt{2}}\bigg[\sqrt{1+\frac{x^{2}}{\lambda}}+\frac{\alpha x}{\sqrt{1+\frac{x^{2}}{\lambda}}}\bigg],
\end{equation}
and
\begin{equation}\label{e2.4}
  \hat{A}^{\dag} = \frac{1}{\sqrt{2}}\bigg[-\sqrt{1+\frac{x^{2}}{\lambda}}+\frac{\big(\alpha-\frac{1}{\lambda}\big) x}{\sqrt{1+\frac{x^{2}}{\lambda}}}\bigg],
\end{equation}
respectively. The operator $\hat{A}$ defined in Eq. (\ref{e2.4'}), satisfies the condition given in Eq. (\ref{cg}), and provides us with the ground state function of $\hat{H}$ as
\begin{equation}\label{gs2}
\varphi_{0}(x)=\mathcal{N}_{0}\bigg(1+\frac{x^{2}}{\lambda}\bigg)^{-\frac{\alpha \lambda}{2}},
\end{equation}
where $\mathcal{N}_{0}$ is the normalization constant. Note that
$\lim_{\lambda\rightarrow \infty}\varphi_{0}(x)=e^{-\frac{\alpha x^{2}}{2}}$, i.e., the ground state of the non-linear harmonic oscillator reduces to the ground state of the linear oscillator when the PDEM no longer remains variable. \\
\noindent The isospectral Hamiltonian $\hat{H}_{-}$ defined in Eq. (\ref{2.2}), becomes
\begin{equation}\label{e2.5}
  \hat{H}_{-}=\frac{1}{2}\bigg[-\bigg(1+\frac{x^{2}}{\lambda}\bigg)\frac{d^{2}}{dx^{2}}-\frac{2x}{\lambda}\frac{d}{dx}
+\frac{\alpha^{2}x^{2}}{1+\frac{x^{2}}{\lambda}}\bigg]-\frac{\alpha}{2}.
\end{equation}
Substitution of Eqs. (\ref{e2.2}) and (\ref{e2.3}) in Eq. (\ref{c}), provides us with the ground state energy of the Hamiltonian $\hat{H}$, as
\begin{equation*}
  E_{0}=\frac{\alpha}{2}.
\end{equation*}
For the present case, the partner Hamiltonian $\hat{H}_{+}$ defined in Eq. (\ref{2.5}), is given as
\begin{equation}\label{e2.6}
  \hat{H}_{+}=\frac{1}{2}\bigg[-\bigg(1+\frac{x^{2}}{\lambda}\bigg)\frac{d^{2}}{dx^{2}}-\frac{2x}{\lambda}\frac{d}{dx}
+\bigg(\alpha-\frac{1}{\lambda}\bigg)^{2}\frac{x^{2}}{1+\frac{x^{2}}{\lambda}}\bigg]-\frac{\alpha}{2}.
\end{equation}
Note that these partner Hamiltonians are related to each other by means of the integrability condition (\ref{2.7}), as
\begin{equation}\label{e2.7}
R(\alpha_{1})=\hat{H}_{+}(\alpha_{1})-\hat{H}_{-}(\alpha_{2})=\alpha_{1}-\frac{1}{\lambda},
\end{equation}
where $\alpha_{1}=\alpha$ and $\alpha_{2}=\alpha-\frac{1}{\lambda}$. By making use of the above information in Eq. (\ref{2.11'}), we obtain the energy spectrum of the given system as
\begin{equation}\label{e2.8}
 E_{n}=\alpha\bigg(n+\frac{1}{2}\bigg)-\frac{1}{2\lambda}n(n+1).
\end{equation}
Note that when $\lambda\rightarrow \infty$, the last relation reduces to the energy spectrum of the standard harmonic oscillator with constant mass. \\
\noindent In order to obtain the excited states of the given system we make use of Eqs. (\ref{e2.4}) and (\ref{gs2}) in Eq. (\ref{2.13}). The first wave function for the system under consideration can be written as
\begin{equation}\label{e2.9}
 \varphi_{1}(x)=\frac{1}{\sqrt{2}}\bigg[-\sqrt{1+\frac{x^{2}}{\lambda}}+\frac{\big(\alpha-\frac{1}{\lambda}\big) x}{\sqrt{1+\frac{x^{2}}{\lambda}}}\bigg]\varphi_{0}(x).
\end{equation}
Let us consider the following identity
\begin{equation*}
 \frac{1}{\sqrt{2}}\bigg[-\sqrt{1+\frac{x^{2}}{\lambda}}+\frac{\big(\alpha-\frac{1}{\lambda}\big) x}{\sqrt{1+\frac{x^{2}}{\lambda}}}\bigg]h_{2}(x)=
\frac{-1}{\sqrt{2}}\bigg[\bigg(1+\frac{x^{2}}{\lambda}\bigg)^{\frac{\alpha \lambda}{2}}\frac{d}{dx}
\bigg(1+\frac{x^{2}}{\lambda}\bigg)^{-\frac{\alpha \lambda}{2}+\frac{1}{2}}\bigg]h_{2}(x),
\end{equation*}
where $h_{2}(x)$ is any arbitrary function. By making use of the above relation in Eq. (\ref{e2.9}), we may rewrite the first wave function in a more simplified form as
\begin{equation}\label{e2.10}
\varphi_{1}(x)=(-1)\mathcal{N}_{1}\bigg(1+\frac{x^{2}}{\lambda}\bigg)^{\frac{-\alpha \lambda}{2}}
\bigg(1+\frac{x^{2}}{\lambda}\bigg)^{\alpha \lambda}\frac{d}{dx}
\bigg(1+\frac{x^{2}}{\lambda}\bigg)^{-\alpha \lambda+1}.
\end{equation}
Similarly we get the next eigenfunction as
\begin{equation}\label{e2.11}
\varphi_{2}(x)=(-1)^{2}\mathcal{N}_{2}\bigg(1+\frac{x^{2}}{\lambda}\bigg)^{\frac{-\alpha \lambda}{2}}
\bigg(1+\frac{x^{2}}{\lambda}\bigg)^{\alpha \lambda}\frac{d^{2}}{dx^{2}}
\bigg(1+\frac{x^{2}}{\lambda}\bigg)^{-\alpha \lambda+2}.
\end{equation}
Repeated application of the above process enables us to obtain the $n^{th}$ wave function of the pertaining systems as
\begin{equation}\label{e2.12}
\varphi_{n}(x)=\mathcal{N}_{n}\bigg(1+\frac{x^{2}}{\lambda}\bigg)^{\frac{-\alpha \lambda}{2}}
\bigg[(-1)^{n}\bigg(1+\frac{x^{2}}{\lambda}\bigg)^{\alpha \lambda}\frac{d^{n}}{dx^{n}}
\bigg(1+\frac{x^{2}}{\lambda}\bigg)^{-\alpha \lambda+n}\bigg],
\end{equation}
where $\mathcal{N}_{n}$ is the normalization constant. \\
\noindent For the sake of simplicity we introduce the dimensionless parameters as
\begin{equation*}
x=\frac{\varsigma}{\sqrt{\alpha}}~~~\mbox{and}~~~\lambda= \frac{\mu}{\alpha},
\end{equation*}
so that the eigenenergies (\ref{e2.8}) and the corresponding wave functions (\ref{e2.12}), are respectively given as
\begin{equation}\label{e23.10'}
 E_{n} = \alpha \bigg[\bigg(n+\frac{1}{2}\bigg) - \frac{1}{2\mu}n(n+1)\bigg],~~~~~~~n=0,1,2,....
\end{equation}
and
\begin{equation}\label{e23.16}
 \varphi_{n} = \mathcal{N}_{n}(-1)^{n}\bigg(1+\frac{\varsigma^{2}}{\mu}\bigg)^{\frac{-\mu}{2}}
               \bigg[\bigg(1+\frac{\varsigma^{2}}{\mu}\bigg)^{\mu}\frac{d^{n}}{d\varsigma^{n}}
               \bigg(1+\frac{\varsigma^{2}}{\mu}\bigg)^{n-\mu}\bigg].
\end{equation}
Note that the term within the parenthesis represents the Rodrigues formula for the $\mu-$dependent Hermite polynomials, since,
\begin{equation*}
 \lim_{\mu\rightarrow \infty}\bigg[(-1)^{n}\bigg(1+\frac{\varsigma^{2}}{\mu}\bigg)^{\mu}\frac{d^{n}}{d\varsigma^{n}}
 \bigg(1+\frac{\varsigma^{2}}{\mu}\bigg)^{n-\mu}\bigg]=(-1)^{n}e^{\varsigma^{2}} \frac{d^{n}}{d\varsigma^{n}}e^{-\varsigma^{2}}.
\end{equation*}
Substitution of last expression in (\ref{e23.16}) yields
\begin{equation}\label{e23.16'}
 \varphi_{n} = \mathcal{N}_{n}\bigg(1+\frac{\varsigma^{2}}{\mu}\bigg)^{\frac{-\mu}{2}}\mathcal{H}_{n}(\varsigma,\mu),
               ~~~n=0,1,2,.....,
\end{equation}
where $\lim_{\mu\rightarrow \infty}\mathcal{H}_{n}(\varsigma,\mu)={H}_{n}(\varsigma)$. \\
Now we investigate certain properties of this class of Hermite polynomials. An appropriate generating function for these $\mu-$dependent Hermite polynomials is given as
\begin{equation}\label{e23.18}
  g(\varsigma,\mu,t)=\bigg(1+\frac{2t\varsigma-t^{2}}{\mu}\bigg)^{\mu-\frac{1}{2}}.
\end{equation}
such that
\begin{equation*}
  \lim_{\mu \rightarrow \infty} g(\varsigma,\mu,t)=e^{2t\varsigma-t^{2}},
\end{equation*}
which is the generating function for the well Hermite polynomials. The generating function defined in Eq. (\ref{e23.18}), can b expressed in terms of the power series as
\begin{equation}\label{e23.19}
\bigg(1+\frac{2t\varsigma-t^{2}}{\mu}\bigg)^{\mu-\frac{1}{2}}=\sum_{k=0}^{\infty}\mathcal{H}_{k}(\varsigma,\mu)\frac{t^{k}}{k!}.
\end{equation}
Explicit expression for the first few polynomials are given as
\begin{eqnarray*}
 \mathcal{H}_{0} &=& 1, \\
  \mathcal{H}_{1} &=& \bigg(2-\frac{1}{\mu}\bigg)\varsigma, \\
 \mathcal{H}_{2} &=& (-1)\bigg(2-\frac{1}{\mu}\bigg)[1-\bigg(2-\frac{3}{\mu}\bigg)\varsigma^{2}], \\
 \mathcal{H}_{3} &=& (-3)\bigg(2-\frac{1}{\mu}\bigg)\bigg(2-\frac{3}{\mu}\bigg)
 \bigg[\varsigma-\frac{1}{3}(2-\frac{5}{\mu})\varsigma^{3}\bigg],\\
 \mathcal{H}_{4} &=& (3)\bigg(2-\frac{1}{\mu}\bigg)\bigg(2-\frac{3}{\mu}\bigg)\bigg[1-2\bigg(2-\frac{5}{\mu}\bigg)\varsigma^{2}
 +\frac{1}{3}\bigg(2-\frac{5}{\mu}\bigg)\bigg(2-\frac{7}{\mu}\bigg)\varsigma^{4}\bigg],\\
\end{eqnarray*}
Note that when the non-linearity parameter $\mu\rightarrow \infty$, all the expressions obtained above, reduce to the well known expressions for the standard Hermite polynomials. Also it is important to remark that similar results can be obtained by solving the system analytically.\\
Let us take the derivative of Eq. (\ref{e23.19}) with respect to $``t"$ and simplify. We finally arrive at
\begin{eqnarray*}
\varsigma\bigg(2-\frac{1}{\mu}\bigg)\sum_{m=0}^{\infty}\mathcal{H}_{m}(\varsigma,\mu)
\frac{t^{m}}{m!}-\bigg(2-\frac{1}{\mu}\bigg)\sum_{m=0}^{\infty}\mathcal{H}_{m}(\varsigma,\mu)\frac{t^{m+1}}{m!}
=\\
\sum_{m=0}^{\infty}\mathcal{H}_{m+1}(\varsigma,\mu)\frac{t^{m}}{m!}+\frac{2\varsigma}{\mu} \sum_{m=0}^{\infty}\mathcal{H}_{m+1}(\varsigma,\mu)\frac{t^{m+1}}{m!}
 -\frac{1}{\mu} \sum_{m=0}^{\infty}\mathcal{H}_{m+1}(\varsigma,\mu)\frac{t^{m+2}}{m!}.
\end{eqnarray*}
This leads to the following expression
\begin{equation}\label{e23.21}
  \mathcal{H}_{m+1}-\varsigma\bigg[\bigg(2-\frac{1}{\mu}\bigg)-\frac{2m}{\mu} \bigg]\mathcal{H}_{m}+m\bigg[\bigg(2-\frac{1}{\mu}\bigg)-\frac{(m-1)}{\mu}\bigg]\mathcal{H}_{m-1}=0,~~~m=0,1,2,....
\end{equation}
Now by differentiating Eq. (\ref{e23.19}) with respect to $``\varsigma"$ and simplifying, we get
\begin{equation*}
  (2-\frac{1}{\mu})\sum_{m=0}^{\infty}\mathcal{H}_{m}(\varsigma,\mu)\frac{t^{m+1}}{m!}=
\sum_{m=0}^{\infty}\mathcal{H}^{'}_{m}(\varsigma,\mu)\frac{t^{m}}{m!}+\frac{2\varsigma}{\mu} \sum_{m=0}^{\infty}\mathcal{H}^{'}_{m}(\varsigma,\mu)\frac{t^{m+1}}{m!}-\frac{1}{\mu} \sum_{m=0}^{\infty}\mathcal{H}^{'}_{m}(\varsigma,\mu)\frac{t^{m+2}}{m!},
\end{equation*}
which leads to the following recursion relation
\begin{equation}\label{e23.20}
  m\bigg(2-\frac{1}{\mu}\bigg)\mathcal{H}_{m-1}=\mathcal{H}^{'}_{m}+2\frac{\varsigma}{\mu} m \mathcal{H}^{'}_{m-1}-\frac{1}{\mu} m(m-1)\mathcal{H}^{'}_{m-2},~~~m=0,1,2,.....
\end{equation}
Note that when $\mu\rightarrow \infty$, the expressions obtained in Eqs. (\ref{e23.21}) and (\ref{e23.20}), become
\begin{equation*}
\mathcal{H}_{m+1}-2\varsigma\mathcal{H}_{m}+2m\mathcal{H}_{m-1}=0,
\end{equation*}
and
\begin{equation*}
  \mathcal{H}^{'}_{m}-2m\mathcal{H}_{m-1}=0,
\end{equation*}
for $m=0,1,2,...$, respectively, i.e., we get back the well known recursion relation for the standard Hermite polynomial.
\subsection{Case 3: $m(x)=\frac{2}{1-(\lambda x)^{2}}$}
\noindent As another example we consider the PDEM of the form
\begin{equation}\label{e3.1}
m(x)=\frac{2}{1-(\lambda x)^{2}},
\end{equation}
such that the Hamiltonian $\hat{H}$ defined in Eq. (\ref{1.2}), of the non-linear oscillator takes the form
\begin{equation}\label{e3.2}
\hat{H}=\frac{1}{4}\bigg[-\bigg(1-(\lambda x)^{2}\bigg)\frac{d^{2}}{dx^{2}}+2\lambda^{2} x\frac{d}{dx}
+4\frac{\alpha^{2}x^{2}}{1-(\lambda x)^{2}}\bigg].
\end{equation}
For the present case, the mass profile encounters a singularity for both positive and negative values of $\lambda$ and our study of dynamics is restricted to the interior of the interval $x^{2}\leq 1/\lambda^{2}$. Thus, the quantum Hamiltonian given in Eq. (\ref{e3.2}), is explicitly Hermitian in the space $L^{2}[-1/\lambda,1/\lambda]$. \\
In order to apply the SUSY QM we consider the super potential of the form
\begin{equation}\label{e3.3}
W(x)=\frac{\alpha x}{\sqrt{1-(\lambda x)^{2}}},
\end{equation}
and the pair of intertwining operators are respectively given as
\begin{equation}\label{e3.4'}
  \hat{A} = \frac{1}{2}\bigg[\sqrt{1-(\lambda x)^{2}}\frac{d}{dx}+\frac{2\alpha x}{\sqrt{1-(\lambda x)^{2}}}\bigg],
\end{equation}
and
\begin{equation}\label{e3.4}
  \hat{A}^{\dag} = \frac{1}{2}\bigg[-\sqrt{1-(\lambda x)^{2}}\frac{d}{dx}+\frac{(2\alpha+\lambda^{2})x}{\sqrt{1-(\lambda x)^{2}}}\bigg].
\end{equation}
Note that $\hat{A}$ defined in Eq. (\ref{e3.4'}), satisfies the constraint (\ref{cg}), and yields the ground state function of the given system as
\begin{equation}\label{gs3}
\varphi_{0}(x)=\mathcal{N}_{0}(1-(\lambda x)^{2})^{\frac{\alpha}{\lambda^{2}}}.
\end{equation}
Also for any differentiable function $h_{3}(x)$, it can be easily shown that
\begin{eqnarray*}
&&\bigg[-\sqrt{1-(\lambda x)^{2}}\frac{d}{dx}+\frac{(2\alpha+\lambda^{2})x}{\sqrt{1-(\lambda x)^{2}}}\bigg]h_{3}(x)=\\
&&(-1)\bigg[(1-(\lambda x)^{2})^{-\frac{\alpha}{\lambda^{2}}}\frac{d}{dx}
(1-(\lambda x)^{2})^{\frac{\alpha}{\lambda^{2}}+\frac{1}{2}}\bigg]h_{3}(x).
\end{eqnarray*}
Thus, the operator given in Eq. (\ref{e3.4}), can be rewritten as
\begin{equation}\label{e3.4n}
  \hat{A}^{\dag} =\frac{-1}{2}\bigg[(1-(\lambda x)^{2})^{-\frac{\alpha}{\lambda^{2}}}\frac{d}{dx}(1-(\lambda x)^{2})^{\frac{\alpha}{\lambda^{2}}+\frac{1}{2}}\bigg].
\end{equation}
For the present case the isospectral Hamiltonians $\hat{H}_{-}$ and $\hat{H}_{+}$ are respectively given as
\begin{equation}\label{e3.5}
  \hat{H}_{-}=\frac{1}{4}\bigg[-\bigg(1-(\lambda x)^{2}\bigg)\frac{d^{2}}{dx^{2}}+2\lambda^{2} x\frac{d}{dx}
+\frac{4\alpha^{2}x^{2}}{1-(\lambda x)^{2}}\bigg]-\frac{\alpha}{2},
\end{equation}
and
\begin{equation}\label{e3.6}
  \hat{H}_{+}=\frac{1}{4}\bigg[-\bigg(1-(\lambda x)^{2}\bigg)\frac{d^{2}}{dx^{2}}
  +2\lambda^{2} x\frac{d}{dx}
+\frac{4\big(\alpha+\frac{\lambda^{2}}{2}\big)^{2}x^{2}}{1-(\lambda x)^{2}}\bigg]+\frac{\alpha}{2}+\frac{\lambda^{2}}{4}.
\end{equation}
In order to determine the ground state energy of the given system we make use of Eqs. (\ref{e3.2}) and (\ref{e3.5}), in Eq. (\ref{c}), and get
\begin{equation}\label{ge3}
  E_{0}=\frac{\alpha}{2}.
\end{equation}
In order to specify the shape invariance condition defined in Eq. (\ref{2.7}), we consider the relations for the partner Hamiltonians $\hat{H}_{-}$ given in Eq. (\ref{e3.5}) and $\hat{H}_{+}$ introduced in Eq. (\ref{e3.6}), such that
\begin{equation}\label{e3.7}
\hat{H}_{+}(\alpha_{1})-\hat{H}_{-}(\alpha_{2})=\alpha_{1}+\frac{\lambda^{2}}{2},
\end{equation}
where the parameters concerning SI are related as
\begin{eqnarray}\label{e3.8}
\alpha_{1}=\alpha,~~
\alpha_{2}=f(\alpha_{1})=\alpha_{1}+\frac{\lambda^{2}}{2}~~\mbox{and}~~ 
R(\alpha_{1})=\alpha_{1}+\frac{\lambda^{2}}{2}.
\end{eqnarray}
In order to determine the energy eigenvalues of the system under consideration we make use of Eqs. (\ref{e3.8}) and (\ref{ge3}) in Eq. (\ref{2.11'}), and get
\begin{equation}\label{e3.9}
 E_{n}=\alpha\bigg(n+\frac{1}{2}\bigg)+\frac{\lambda^{2}}{4}n(n+1).
\end{equation}
By using Eqs. (\ref{e3.4n}) and (\ref{gs3}) in recurrence relation given in Eq. (\ref{2.13}), we can obtain the corresponding wave functions as
\begin{equation}\label{e3.10}
\varphi_{n}(\varrho)=\mathcal{N}_{n}[1-(\upsilon \varrho)^{2}]^{\frac{1}{2\upsilon^{2}}}
\bigg[(-1)^{n}[1-(\upsilon \varrho)^{2}]^{-\frac{1}{\upsilon^{2}}}\frac{d^{n}}{d\varrho^{n}}
[1-(\upsilon \varrho)^{2}]^{\frac{1}{\upsilon^{2}}+n}\bigg],
\end{equation}
where $\varrho=x\sqrt{2\alpha}$ and $\upsilon=\lambda/\sqrt{2\alpha}$, are the dimensionless variables and $\mathcal{N}_{n}$ is the normalization constant. Note that the term within the parenthesis represents the Rodrigues formula of the modified Hermite polynomials.\\
The appropriate generating function for the system under consideration is given as
\begin{equation}\label{3.18}
  g(\varrho,\upsilon,t)=[1-\upsilon^{2}(2t\varrho-t^{2})]^{\frac{-1}{\upsilon^{2}}-\frac{1}{2}}.
\end{equation}
Note that this choice of generating function satisfies the correct limit defined above. Due to the existence of this generating function we can obtain the recurrence relations for the modified Hermite polynomials in parallel to the standard Hermite polynomials. The generating function defined in Eq. (\ref{3.18}), in terms of power series is expressed as
\begin{equation}\label{3.19}
 [1-\upsilon^{2}(2t\varrho-t^{2})]^{\frac{-1}{\upsilon^{2}}-\frac{1}{2}}
 =\sum_{k=0}^{\infty}\mathcal{H}_{k}(\varrho,\upsilon)\frac{t^{k}}{k!}.
\end{equation}
Expanding the L.H.S of the above expression and equating the coefficients of powers of $t$, we obtain explicit expressions for first few modified Hermite polynomials as
\begin{eqnarray*}
 \mathcal{H}_{0} &=& 1, \\
  \mathcal{H}_{1} &=& (2+\upsilon^{2})\varrho, \\
 \mathcal{H}_{2} &=& (-1)(2+\upsilon^{2})[1-(2+3\upsilon^{2})\varrho^{2}], \\
 \mathcal{H}_{3} &=& (-3)(2+\upsilon^{2})(2+3\upsilon^{2})\bigg[\varrho-\frac{1}{3}(2+5\upsilon^{2})\varrho^{3}\bigg],\\
 \mathcal{H}_{4} &=& (3)(2+\upsilon^{2})(2+3\upsilon^{2})\bigg[1-2(2+5\upsilon^{2})\varrho^{2}
 +\frac{1}{3}(2+\upsilon^{2})(7\tilde{\lambda}-2)\varrho^{4}\bigg],\\
 \mathcal{H}_{5} &=& (15)(2+\upsilon^{2})(2+\upsilon^{2})(2+\upsilon^{2})
 \bigg[\varrho-\frac{2}{3}(2+\upsilon^{2})\varrho^{3}+\frac{1}{15}(2+\upsilon^{2})
 (2+\upsilon^{2})\varrho^{5}\bigg].
 \end{eqnarray*}
Note that in the harmonic limit $\upsilon \rightarrow 0$, all the results obtained for the system under consideration reduce to the well known results of the celebrated harmonic oscillator. The recurrence relations for this family of Hermite polynomials can be determined in the similar way as done for the last two cases.
\section{Conclusion}
\noindent In the present work we have first discussed the problem of ordering ambiguity in the kinetic energy term that arises due to the variable mass and then quantized our system by following the recipe given by L\'{e}vy-Leblond \cite{levy1995position}, originally proposed by Von Roos \cite{von1983position}. It is important to remark that the quantum Hamiltonian, obtained is manifestly Hermitian. Furthermore, we have provided a general recipe for obtaining the algebraic solutions of the quantum systems with PDEM. This formalism is based on the integrated concepts of SUSY QM and the beautiful property of shape invariance. It is worth mentioning over here that we have restricted ourselves to the study of the PDEM quantum systems with shape invariant potentials that belong to the translation class.\\
\noindent For the sake of illustration we have chosen a class of non-linear oscillators with spatially varying mass. In each case, explicit expressions for the energy spectrum and the corresponding wave functions in terms of the modified Hermite polynomials has been obtained. The wave functions belongs to the family of the orthogonal polynomials that can be considered as the modification of the standard Hermite polynomials. It has been shown that under the harmonic limit, the results obtained for PDEM quantum systems reduce to the corresponding results for the harmonic oscillator with constant mass.

\end{document}